\documentclass{article}

\usepackage{PRIMEarxiv}

\usepackage[utf8]{inputenc} 
\usepackage[T1]{fontenc}    
\usepackage{hyperref}       
\usepackage{url}            
\usepackage{booktabs}       
\usepackage{amsfonts}       
\usepackage{nicefrac}       
\usepackage{microtype}      
\usepackage{lipsum}
\usepackage{fancyhdr}       
\usepackage{graphicx}       
\graphicspath{{media/}}     
\usepackage{longtable}
\usepackage{ragged2e}
\title{Serious Games: An Updated Systematic Literature Review
\thanks{\textit{\underline{Citation}}: 
\textbf{Authors. Title. Pages.... DOI:000000/11111.}} 
}

\author{
  Shuja Ud Din \\
  Faculty of Computing\\ 
  Riphah International University\\
  Islamabad, Pakistan \\
  \texttt{qazi.shujauddin@riphah.edu.pk} \\
  \And
  Muhammad Zeeshan Baig\\
  Academy of Interactive Entertainment (AIE) Institute\\
Sydney, Australia\\
\texttt{Zeeshan.baig@aieinstitute.edu.au}\\
   \And
  Muhammad Khateeb Khan \\
  Faculty of Computing \\
  IQRA University Islamabad Campus (IUIC)\\
  Pakistan \\
  \texttt{muhammad.khateeb@iqraisb.edu.pk} \\
}

\begin{document}
\maketitle

\begin{abstract}
Serious games are simulation software designed to assist people in learning the practical concepts of various application fields such as Health, wellness, Education and Culture. People improve their individual knowledge, skills and attitude through training. This study identified the changing trends with existing studied applications, approaches and methods. We collected 37 papers from Google Scholar, Elsevier, Springer, IEEE Xplore and ACM Digital library. We have collected the evidence answer to six research questions and analyzed the result and identify the change in trends with the comparison with previous systematic literature review (SLR) results. We achieved the best results by techniques(questionnaires and interviews) and procedure (pre$/$post). Our findings will be useful for practitioners and researchers who can test serious games in different fields.
\end{abstract}

\keywords{Serious Games \and Evaluation \and Game-Based Learning \and Learning Analytics}

\section{Introduction and Motivation}
\label{intro}
\cite{1} Introduced the concept of serious games. Serious games consider only the educational intent, not the amusement aspect. Initially, serious games were based on card and board games. Now, it's becoming the trending technique that refers to forms of video game-based learning and training in various fields, including marketing, military, medicine, business, and education. Serious games utilize formal and informal techniques and focus on all-ages audiences \cite{2}. Serious games are developed for learning in a sterile environment, different from simulation, training, learning, testing, and diagnostics. Nowadays, serious games are used in healthcare, communication, defense, education, culture, and politics.

Serious games are active and entertaining learning environments developed for various digital platforms such as computers and smartphones. Learning in serious games mostly takes place through gameplay, where learners must complete the challenges using their in-game skills. Subjective experiences which seek the skills of learners are defined as Challenges. The challenges include the learner cooperating, competing, experimenting, and exploring with other learners. DE U (Usability), DE C (Cognitive behavior), and DE P (Playability) are the common combinations of design elements (DEs) and challenges used in serious games  \cite{8}.

Serious games are designed purely for learning purposes. It provides learning assistance in many domains, such as health, wellness, support, social, education, professional learning, and training. This research gathers information about serious games to find the answers to research questions related to the development, implementation, evaluation, and application of serious games. We have performed a literature review highlighting current research trends in serious games.

The objective of this study is as follows: 
\begin{itemize}
  \item Classify the application areas where there have been significant evaluations of the games.
  \item Find out different types of assessment tests used for serious games 
  \item Collect and evaluate current processes, approaches, and strategies for analyzing serious games.
  \item Identify and analyze the latest trends and research in serious games \cite{11}.
\end{itemize}

\section{Related Work}

Serious games are simulations designed to assist people in learning practical concepts through training. It has been used in health, wellness, education, and culture to improve knowledge, skills, and attitude. Numerous studies have shown the effectiveness of Serious games in improving the skills and attitudes of individuals. In this paper, we presented an analysis of the various aspect of Serious games in terms of usability, effectiveness, and applicability.

McCallum and Boletsis \cite{12} published a study that described dementia disease and its symptoms (Alzheimer's disease (AD) and Mild Cognitive Impairment (MCI)). For this purpose, they gathered the Serious games related to dementia disease and its symptoms, i.e., Smart Brain Games, Big Brain, Wii Fit, Academy Lumosity, and Wii  Sports. These Serious games were totally about dementia disease, which was uncharted. Furthermore, the researchers performed a literature review to gather games and conducted an evaluation test on patients with dementia to determine the effect of transferable and long-lasting daily activities. 

Drummond et al. \cite{13} performed a Systematic Literature Review (SLR) of serious games articles to guide the public and patients about asthma. They evaluated the results on the patient's behavior, knowledge, and clinical outcomes about asthma. Furthermore, the SLR was performed from March 1980 to December 2015 and selected 12 articles related to Serious games about asthma education from online databases, i.e., Web of Science, Embase, Cochrane Library, Psych Info, and PubMed Science. They have selected 12 studies and ten Serious games for the analysis and eight serious games out of ten related to children with asthma. The majority of the Serious games were related to improving the satisfaction level and knowledge of asthma in children. Some studies assess serious games'  effect on clinical outcomes, but the studies have not found significant results. They concluded that Serious games used for asthma education to improve patients' and children's satisfaction and knowledge have less or no improvement in clinical results and behaviors.

A systematic literature review on usability evaluation methods of serious games has been performed by Gomez et al. \cite{14}. For this reason, the six research questions answered related to the evaluation of usability methods. Furthermore, the researchers selected 187 articles from various online databases such as IEEE-Xplore, ACM Library, and ISI Web of Knowledge. The conclusion of this SLR was a usability evaluation of these Serious games and explained the application of focal trends. 

The study in \cite{15} performed a systematic analysis of the literature on Serious games and computer games. The aim is to analyze the positive impact of Serious games on teenagers above 14 years old in learning, skills development, and interaction. The author selected 129 papers from different online databases, including ASSIA, Science Direct, Biomed Central, and ACM. The analyses on these games had positive learning outcomes and highlighted difficulties in learning outcomes. Similarly, the study motivates the students to use these games to improve their skills and knowledge.

In another study, Wang et al. \cite{16} performed a systematic analysis of the literature under the PRISMA guideline. The research aim was to identify the current trends of Serious games in the training of health care professionals, especially games that mainly show game assessment and developmental methodologies. Furthermore, the researchers selected about 48 articles in different online databases such as Cochrane, EMBAE, and PubMed, from December 2007 to 2014. In 48 papers, the study identified 42 unique serious games. The study concluded that serious games are widely used in health care training and obtain several learning objectives.

The study \cite{57} performed a systematic review of the literature
about serious game usage in science education from 2016 to 2020. For this purpose, 39 articles included in electronic databases Web of Sciences targeted list journal and conference are Arts $\&$ Humanities Citation Index (A$\&$ HCI), Science Citation Index- Expanded (SCI-Expanded) and Social Science Citation Index (SSCI).

The study \cite{58} performed a systematic review of the literature on serious games in the area of Business Process Management (BPM). The research objective was to analyze and explain the 15 BPM games based on their learning objectives, aesthetics, technologies, and mechanics. For this purpose, 15 papers are extracted and included from the electronic databases Scopus and Web of Sciences from 2000 to 2020.

A study similar to \cite{11} performed a systematic review of the literature on the assessment of serious games. The research objective is to evaluate serious games through different domains, methods, approaches, and assessment methods. Furthermore, the author selected 102 articles from various online databases to answer the six research questions. The study concluded that answering six research questions had identified different application areas, serious game types, procedures, quality attributes, and methods for assessing Serious games. 

\section{Methodology}
\subsubsection{Protocol}
The systemic literature review process includes defining, reviewing, and analyzing all existing reteach on the specific research area, topic, or subject. The extracted articles, in systematic literature, were named primary studies, and systematic evaluation is a type of secondary research \cite{10}.

This research was carried out as a systematic analysis of literature based on initial guiding principles suggested by \cite{18} which are one of the most commonly known and generally recognized in software engineering. The proposed guidelines have 3 stages which define below:

\begin{table}[htbp]
\caption{Proposed Guideline Stages }
\begin{center}
\begin{tabular}{|c|c|c|}
\hline
 Sr. No & Stages Name & Steps in each stage \\ 
\hline \hline
1&  Planning the  review & 1. Identification of the need for SLR\\
\cline{3-2}
& & 2. Developing protocol\\
\cline{3-2}
& & 3. Validate protocol\\
\hline
2&  Conducting the review & 1. Identification of research\\
\cline{3-2}
& & 2. Selection of primary studies\\
\cline{3-2}
& & 3. Study quality assessment \\
\cline{3-2}
& & 4. Data extraction and monitoring\\
\cline{3-2}
& & 5. Data Synthesis \\
\hline
3&  Reporting the Review  & 1. Specifying mechanisms of dissemination\\
\cline{3-2}
& & 2. Formatting the main report\\
\cline{3-2}
& & 3. Evaluating the Report\\
\hline
\end{tabular}
\end{center}
\end{table}

\subsection{Research questions}
To accomplish the research objectives, six research questions were established. These questions help to gather all the details necessary to assess the various evaluations. The addressed research questions in this research are:

\begin{itemize}
\item RQ1: What are the application domains in which serious games have been assessed?
\item RQ2: What are the types of serious games that have been assessed within the former domains?
\item RQ3: What methods, techniques, and quality models have been used to assess these serious games?
\item RQ4: What are the characteristics of the quality models that have been assessed?
\item RQ5: How the evaluation models, techniques, or methods are applied to assess a serious game?
\item RQ6: What is the size of the population involved in the existing assessment experiences of the serious games?
\end{itemize}

The above-mentioned research questions can be categorized into two areas of interest. RQ1 and RQ2 are used to evaluate the series of games and their involved features. These questions highlighted the application domain, knowledge area, and the type of serious games which were assessed by the various evaluation methods.

In comparison RQ3, 4, 5, and 6 concentrate on the techniques of assessment and its features. To answer research question 3 we studied the assessment processes used to test serious games and we examine the models or structures underlying these processes. For RQ4, we examined the various characteristics evaluated by researchers in their reviews of serious games. About RQ5, we defined the techniques followed by researchers in an assessment period of serious games. Finally, RQ6 indicates the population size that participates in the serious games assessment session [8].

\subsection{Search Strategy}
The search strategy's goal is to classify the primary studies. A systematic review has been conducted to answer the above-mentioned research questions.

Next, we have considered the search keywords. Therefore, generic terms were used to ensure that many of the related scholarly articles were included in the study. “Serious game" and "Evaluation" were the key search words. The following search string was developed and mentions the step in Kitchenham et al.~\cite{18}.

\begin{itemize}
\item Keywords had extracted from the queries by defining the core concepts.
\item Classify keywords with different synonyms and spellings.
\item Review the keywords that we have in any related articles.
\item Boolean OR use to include different spelling and synonyms
\item To connect keywords using Boolean AND.
\end{itemize}

We did some primary searches to test the search string and refined it. The resulting search strings had accompanied by the Boolean expression ‘‘(A1 OR A2 OR A3) AND (B1 OR (B2 AND (C1 OR C2 OR C3)))” following table again represented the search keywords [8].

\begin{table}[htbp]
\caption{Search Keywords}
\begin{center}
\begin{tabular}{|c|c|c|c|}
\hline
A1. Evaluation & B1. Serious Game & C1. Education\\
\hline
A2. Validation & B2. Simulation Game & C2. Teaching\\
\hline
A3. Assessment &  & C3. Training\\
\hline
\end{tabular}
\label{tab1}
\end{center}
\end{table}

The search had carried out in online databases which were: IEEE Xplore, Science Direct (Elsevier), Springer Link, Google Scholar, and ACM Digital Library. The tool which we have used to gather search information is MS Excel.
\subsection{Criteria of inclusion and exclusion and study selection}
\textbf{Stage 1:}
Initially, we selected the research articles by analyzing the title and abstract of every research article found by the search string. Each paper's title and abstract were studied according to the inclusion and exclusion criterion where we excluded the irrelevant papers. Throughout this selection process step, the identified research papers were marked as Non- Selected or Selected papers.

\textbf{Stage 2:}
The research papers which were selected in stage 1 were further filtered out after the complete analysis of each paper. We observed the papers were relevant to “serious games” but unable to answer the defined research questions. We excluded all these papers and considered only such studies which were able to provide the answer to every research question. The complete detail of followed inclusion and exclusion criteria is mentioned in the subsequent section.

\begin{table}[htbp]
\caption{Inclusion/Exclusion criteria}
\begin{center}
\begin{tabular}{|p{3cm}|p{12cm}|}
\hline
Inclusion criteria & 1 Select the articles which depict techniques, models, or strategies to assess the serious games.\\
& 2 Papers that describe the evaluation procedure of serious games.\\
& 3 Papers show case studies with relevant applications to evaluate the techniques, models, or strategies for serious games.\\
\hline
Exclusion criteria & 1 Research whose primary aim is not to assess methods for serious games. \\
& 2 Papers that indicate a serious game, and don't include detail on its evaluation.\\
& 3 Articles that show the findings of evaluating a serious game but don't provide any detail about the methodology used for assessment.  \\
& 4 Not included those articles whose models of evaluation are out from the perspective of serious games.  \\
& 5 Publications are excluded if only the abstract is available but not the complete text.  \\
& 6 The article which is written in other languages except English will be omitted.\\
& 7 Identical articles (same document taken from various databases).\\
& 8 Identical documents of the same study (In the research, the most reliable version of the report was used because various copies of the report occur in separate databases).\\
\hline
\end{tabular}
\label{tab1}
\end{center}
\end{table}

\subsection{Quality Assessment}
During the data extraction process, the quality of each paper was evaluated. For each relevant paper, a questionnaire was filled out, which was defined as a quality instrument. The questionnaire of the assessment contains nine questions of quality assessment (QA) which consist of two parts. The first part contains questions to find the paper quality about the main topic of the literature (QA1-QA4) whereas; the objective of the second part is to find the quality of paper information to know about the paper's relevance with SLR (QA5-QA9). The nine quality assessment (QA) questions that have been used are described as follows:

\begin{itemize}
    \item Q1. Does the article demonstrate a method of evaluation? 
    \item Q2. Does the article reveal details regarding the technique used for evaluation?
    \item Q3. Does the article provide model information that supports the method of evaluation? 
    \item Q4. Is the article using examples to illustrate the application of the assessment method?
    \item Q5. Is the paper requiring the relevant details to be found to address RQ1?
    \item Q6. Does the article require RQ4 to be answered?
    \item Q7. Is the article require the necessary details to address RQ5?
    \item Q8. Is the article enabling the correct details to satisfy RQ6?
     \item Q9. Is the article allowing the relevant information to address in RQ2?
\end{itemize}
\begin{figure}[htbp]
\centerline{\includegraphics[width=9cm, height=5cm]{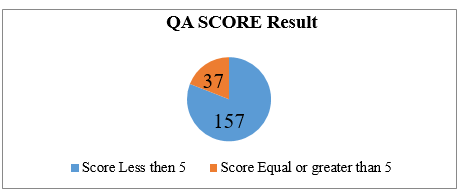}}
\caption{Quality assessment Percentage score}
\label{QA Result}
\end{figure}
Figure \ref{QA Result} indicates the percentage for quality evaluation. We have included 37 papers in the synthesis process in the primary study, whose quality assessment score was equal to or greater than 5. We have implemented the third step and concluded 194 articles. We applied a quality assurance method in that step, and articles that scored less than 5 were left out.

\section{Results}
\subsection{Selection Process}
\begin{figure}[htbp]
\centerline{\includegraphics[width=8cm, height=6cm]{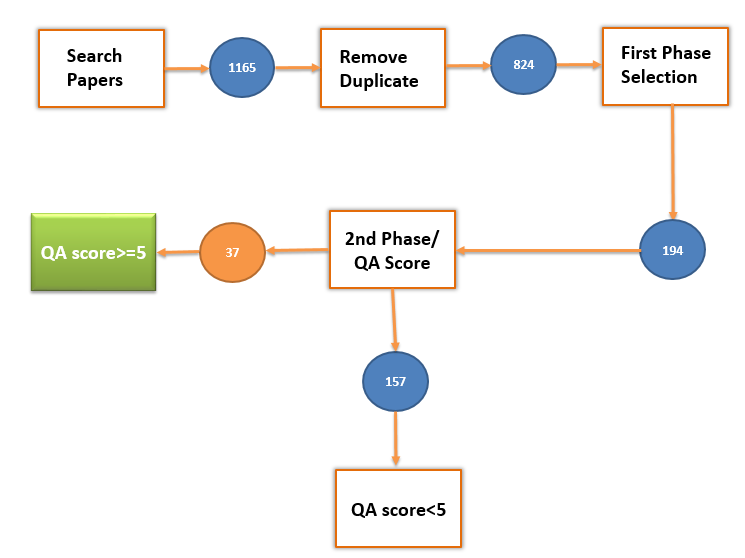}}
\caption{Selected Primary Studies from Online Databases }
\label{QAProcess}
\end{figure}

 As the screening process started shown in Fig \ref{QAProcess} and found a total of 1165 research articles from different electronic databases. Our process was completed at two different levels, in our first phase we removed the duplicate articles and shortlisted 824 articles for further processing. In the next level, we studied the title and the abstract of each article and based on the inclusion and exclusion criteria selected only 194 articles that satisfy the criteria. We studied the selected articles in detail and finally, we shortlisted 37 articles that have prominent relevancy and closeness to the required criteria based on QA score. 

\subsection{Discussions}
\textbf{RQ1. What are the application areas where serious games were evaluated?}

We identified six major application areas by applying the inclusion and exclusion criteria of 37 selected papers to answer this question. The detailed information on the above application areas is given in Table (\ref{RQ1}).

\begin{table}[htbp]
\caption{Categories of Application fields}
\begin{center}
\begin{tabular}{| p{3cm} | p{7cm}| p{7em} |}
\hline
\textbf{Categories}& \textbf{Definitions} & \textbf{Primary Studies} \\ 
\hline
Culture & This domain covers serious games for cultural training ~\cite{11}~\cite{19}. &~\cite{19}\\
\hline
Health and Wellness & Created these serious games include application domain related to improving the quality of health life of the people and sharing good habit knowledge during their everyday life ~\cite{11}. & ~\cite{20,22,29,30,31,32,37,40,42,47,48} \\
\hline
Education & This application category contains serious games for teaching, supporting, evaluating, and inspiring students, and awareness in various areas of formal education ~\cite{11}. & ~\cite{21,23,24,25,26,27,28,33,34,35,36,41,43,44,45,49,51,52,53,54,55}\\
\hline
Support & Developed the application field involves serious games that encourage and assist people in decision-taking in life~\cite{11}. &~\cite{38}\\
\hline
Professional learning and training & This application area involve serious games that have been used to train and teach their workers in companies~\cite{11}. &~\cite{39}\\
\hline
Social & This area of application has been involved in serious games for training social skills~\cite{11}. &~\cite{46,50}\\
\hline
\end{tabular}
\label{RQ1}
\end{center}
\end{table}

\begin{figure}[htbp]
\centerline{\includegraphics[width=9cm, height=5cm]{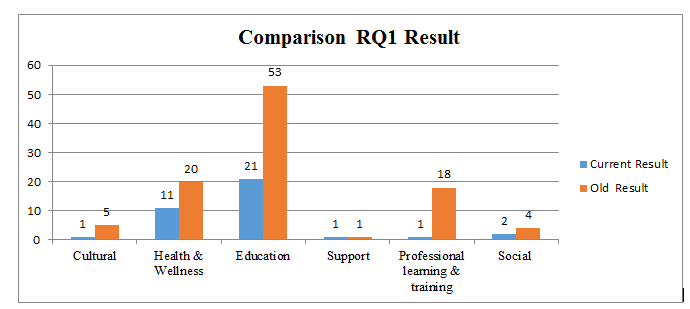}}
\caption{Comparison Results RQ1 with Old Findings}
\label{CompRQI}
\end{figure}

Fig (\ref{CompRQI}) showed the comparison with the previous SLR ~\cite{11}. In  ~\cite{11}  education domain was depicted as the most prominent and top domain which assessed serious games. Similarly, the education domain remained at the top in our new findings, as given in Fig (\ref{CompRQI}). Health and Wellness have remained the 2nd top category in our old and newly conducted literature review. Few serious games are evaluated n the domain of social and cultural categories in both our old and latest SLR results. We found a few serious games in the categories of professional learning and training in our new findings, it was the 4th highest category among all six categories in old SLR ~\cite{11} results. 

\textbf{RQ2. What kinds of serious games were evaluated within the prior domains?}

To evaluate the question, we identified 8 major types of serious games by applying inclusion and exclusion criteria to 37 selected articles and the results are given in Table (\ref{RQ2}).
\begin{table}[htbp]
\caption{Types of Serious Games}
\begin{center}
\begin{tabular}{| p{2cm} | p{7cm}| p{2cm} |}
\hline
\textbf{Types of  games}& \textbf{Definitions} & \textbf{Primary Studies} \\ 
\hline
Computer games & Developed serious games as software for computers~\cite{11}. &~\cite{19,20,21,27,31,32,36,40,47,48}\\
\hline
Virtual worlds & Created serious games as a virtual 3D interactive environment.&~\cite{22,28,29,39,44,46,50,51,52,54,55} \\
\hline
Lego-based Games & Serious games are created as computer software which utilizes as a part of the game Lego construction toys ~\cite{11}. &~\cite{23}\\
\hline
Video games &Serious games playable on certain platforms, such as consoles for video games (Xbox, PlayStation, etc.)~\cite{11}. &~\cite{25,30,33,34,42,43,45}\\
\hline
Board game & Created serious games as a board game~\cite{11}. &~\cite{26,28}\\
\hline
Web-based game & Created serious games as a web app~\cite{11}. &~\cite{35,41,53}\\
\hline
Mobile game & Created serious game as a mobile app~\cite{11}. &~\cite{37,49}\\
\hline
MMORPG & Serious games created as Online Role-Playing Massively
Multiplayer game~\cite{11}. & No study found.\\
\hline
\end{tabular}
\label{RQ2}
\end{center}
\end{table}

\begin{figure}[htbp]
\centerline{\includegraphics[width=9cm, height=5cm]{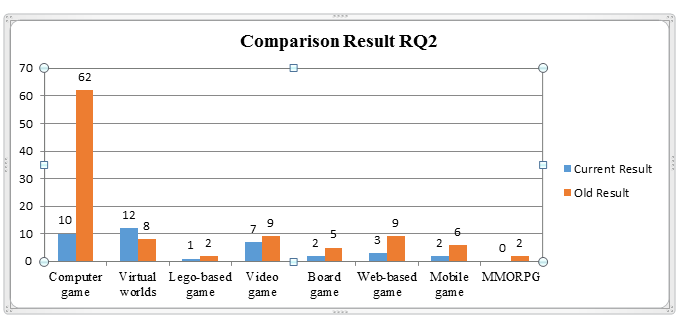}}
\caption{Comparison Results RQ2 with Old Findings}
\label{CompRQ2}
\end{figure}

 Results of a previous SLR~\cite{11} are presented in Fig (\ref{CompRQ2}), where game types evaluate all serious games. According to the results, “Computer Games” is categorized as the highest game type, whereas it lies at the top second place according to the current study. “Virtual World Games (3D)” have emerged as the most assessed game type in serious games. According to our research findings, such games are beneficial in learning the concepts of respective domains. Previously, “Virtual World Games (3D)” was a newborn game type in serious games and very few studies were available that evaluated this type. Similarly, “Video Games” is found to be the 3rd highest game type that was previously not that popular in serious games, and very few studies were found that evaluated this game type as shown in Fig (\ref{RQ2}).    
 On further review, the serious games ( Lego-based and MMORPG types) were few in number in previously conducted SLR while in the updated SLR there was no evidence of such game types.

\textbf{RQ3. Which quality model, approaches, strategies, techniques, or methods were used in evaluating these serious games?}

\begin{figure}[htbp]
\centerline{\includegraphics[width=9cm, height=5cm]{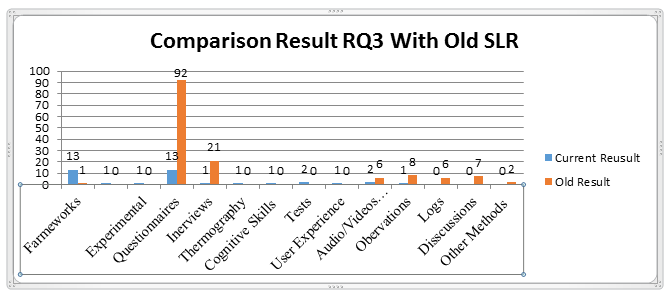}}
\caption{Comparison Results RQ3 with Old Findings}
\label{compRQ3}
\end{figure}

We identified the quality model, approaches, and techniques to evaluate the serious games to analyze the above question.  The table (\ref{RQ3}) describes the approaches and techniques to assess serious games in both previous SLR  and new findings.
\begin{table}[htbp]
\caption{Assessment Techniques}
\begin{center}
\begin{tabular}{| p{2cm} | p{7cm}| p{2cm} |}
\hline
\textbf{Technique}& \textbf{Definition} & \textbf{Primary Studies} \\ 
\hline
Frameworks & The procedure utilizes a set of methods and steps that the authors have described evaluating the serious game~\cite{11}. &~\cite{19,23,25,30,33,34,36,38,41,45,49}\\
\hline
Questionnaires & The procedure follows question forms to evaluate the serious game~\cite{11}.&~\cite{24,35,39,42,43,44,48,50,47,53,54,22}\\
\hline
Observations & The approach is focused on observing the meeting progress to evaluate the serious game~\cite{11}. & ~\cite{55}\\
\hline
Audio/Videos Feedback &The method uses audio, and video feedback from experts to evaluate the serious game~\cite{11}. &~\cite{52,47}\\
\hline
Tests & The method used to perform tests to assess the serious games. & ~\cite{32,46}\\
\hline
Evaluation and Training & The method uses evaluation. & ~\cite{20}\\
\hline
\end{tabular}
\label{RQ3}
\end{center}
\end{table}
 Fig (\ref{compRQ3}) shows the trends and evaluation models or approaches used to assess serious games. Questioners and frameworks had been the top method for evaluating serious games. In previous SLR results, “questioners” was the topmost evaluation method to assess the most serious games, whereas only one research study used the “framework” evaluation method to evaluate the serious games. The evaluation methods named (“Evaluation and Tanning,” “Experimental,” ”Thermography,” “Cognitive Skills,” “Tests,” and “User experiences,”) were new approaches identified in the current SLR whereas previous studies have found no result in these evaluation models which have been shown in Fig (\ref{compRQ3}). The evaluation model named “Logos,” ”discussions,” and “other methods” these approaches had been identified in previous findings, whereas found no result in the updated SLR.

\textbf{RQ4 What are the qualitative attributes that were evaluated?}
\par
In, RQ4 discusses the software quality attribute to evaluate the. Serious games. The quality attributes are those attributes that measure how well serious games have been designed.
The purpose of the research question is to determine the quality attributes that the researchers have taken into the record to assess their serious games and found 13 different quality attributes to analyze serious games as given in Table (\ref{RQ4}).

\begin{table}[htbp]
\caption{Quality Attributes}
\begin{center}
\begin{tabular}{| p{3cm} | p{7cm}| p{2cm} |}
\hline
\textbf{Quality attributes}& \textbf{Definition} & \textbf{Primary Studies} \\ 
\hline
Game design & Design serious games, also making aesthetic of serious games ~\cite{11}. &~\cite{19,20}\\
\hline
Performance & To measure the performance of the serious game ~\cite{11}.& ~\cite{20,32,34,39,47,53}\\
\hline
Learning outcomes & What the learners should think as a consequence of playing a serious game ~\cite{11}. &~\cite{21,44}\\
\hline
User interface &The user's interaction with serious game. &~\cite{22}\\
\hline
Usability & Serious games assess how easy the user interface is to use. & ~\cite{23,33,37,40,42,43,48,51}\\
\hline
User's satisfaction & User attitude to the serious game~\cite{11}. &~\cite{24}\\
\hline
Cognitive behavior & The potential of the serious game to create cognitive effects on users' behavior. &~\cite{26,30,31}\\
\hline
The user’s experience & The emotion, attitude, and actions of the user are using the serious game. & ~\cite{27}\\
\hline
Efficacy & The serious game ability to deliver the desired result. &~\cite{28,29,36,38}\\
\hline
Learning Efficacy & The serious game ability to consider the robustness of the study evidence in support of enhanced student performance by implementing a particular teaching technique. & ~\cite{35,46,54,55}\\
\hline
Learnability & Measures how rapidly and easily users learn the basic functionality of serious games. &~\cite{43}\\
\hline
Pedagogical aspects & Educational features of the serious games ~\cite{11}. &~\cite{49,52}\\
\hline
Playability & The serious game ability to being played~\cite{11}. & ~\cite{50}\\
\hline
\end{tabular}
\label{RQ3}
\end{center}
\end{table}

\begin{figure}[htbp]
\centerline{\includegraphics[width=9cm, height=5cm]{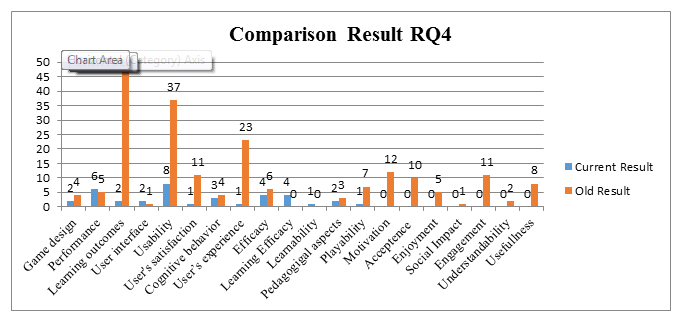}}
\caption{Comparison Results RQ4 with Old Findings}
\label{compRQ4}
\end{figure}

As shown in Fig (\ref{compRQ4}) the comparison of quality attributes with the previous SLR~\cite{11} results. In updated results, most of the serious games have identified quality attributes usability while in the previous results \cite{11} usability was the second-highest category. In the previous result, most studies used the quality attributes named learning outcomes to evaluate serious games. The quality attribute named Performance, Efficacy, Game Design, User Interface, user satisfaction, user experiences, pedagogical aspects, and Learning efficacy were evaluated in different serious games domain in new findings. There were few amounts of quality attributes mentioned in previous studies.

\textbf{RQ5 How are the assessment methods, techniques, or modes used for evaluating a serious game?}
In response to the above question, we have found various stages to accomplish serious game assessments. We identified the 15 types of procedures in this study to evaluate the serious games as shown in Table (\ref{RQ5}).

\begin{table}[htbp]
\caption{Procedure Types.}
\begin{center}
\begin{tabular}{| p{8em} | p{5cm} |}
\hline
\textbf{Procedure }& \textbf{Primary Studies} \\ 
\hline
Pre-Test Questionnaire  & ~\cite{29,43}\\
\hline
Pre/Post &~\cite{20,21,22,26,29,32,38,39,40,41,50,51,52,55,36} \\
\hline
Usability Testing  & ~\cite{27,24,28}\\
\hline
End User Testing & ~\cite{25}\\
\hline
Post-Hoc Tests &~\cite{28,33}\\
\hline
Preliminary Tests&~\cite{30}\\
\hline
Experimental Test  &~\cite{31,34}\\
\hline
Mann-Whitney TESTS & ~\cite{35}\\
\hline
User Testing & ~\cite{37}\\
\hline
Shapiro–Wilk test &~\cite{37}.\\
\hline
Post-Test &~\cite{46}\\
\hline
Five Dimension Evaluation Framework &~\cite{23}\\
\hline
\end{tabular}
\label{RQ5}
\end{center}
\end{table}

\begin{figure}[htbp]
\centerline{\includegraphics[width=9cm, height=5cm]{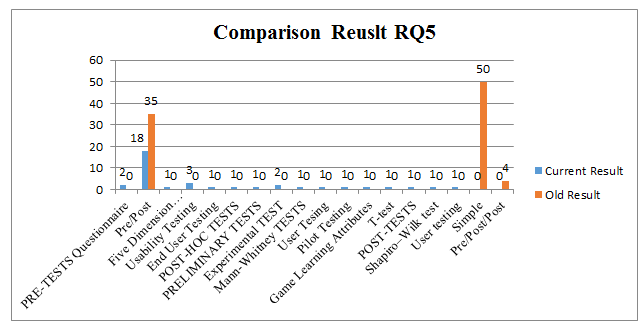}}
\caption{Comparison Results RQ5 with Old Findings}
\label{compRQ5}
\end{figure}

All evaluation models were applied to evaluate the serious game as given in Fig (\ref{compRQ5}). In previous SLRs Simple, Pre/Post, and Pre/Post/Post approaches were used to assess the serious games while in the current SLR Pre/Post approach was used to assess the serious games. A Simple technique commonly used to evaluate serious games was the highest category in previous findings. Pre/Post was the 2nd highest category in the old result, and almost 40 percent of the studies used the approach to evaluate the serious game. In new SLR results, 48$\%$ of the research articles used Pre/Post to analyze serious games and while the rest of 52$\%$ was analyzed through newly identified methods as shown in Fig(\ref{compRQ5}).

\textbf{RQ6. What is the population size involved in the serious game's existing evaluation experiences?}

To analyze this question, we found the size of the participant groups involved in evaluating the serious game. We identified these groups after analyzing procedures, techniques, and quality characteristics assessed in the primary studies. In the analysis of primary studies, 70 $\%$ of the population size is less than or equal to 50 people, and 5$\%$ of the population size consists of greater than 50 or less than 100 people. The remaining populations are shown in Table(\ref{RQ6}).

\begin{table}[htbp]
\caption{Population ranges identified in Primary research.}
\begin{center}
\begin{tabular}{| p{8em} | p{5cm} |}
\hline
\textbf{Range}& \textbf{Primary Studies} \\ 
\hline
[1,10]  & ~\cite{29,43}[29], [43]\\
\hline
Pre/Post & ~\cite{20,29,30,31,33,40,48,49}\\
\hline
[11,20]  & ~\cite{23,25,26,38,46}\\
\hline
[21,30] &~\cite{19,28,37,4243,50,55} \\
\hline
[31,40] &~\cite{21,24,27,35,44,47,55}\\
\hline
[41,50]&~\cite{32,34,36,52}[32], [34],[36],[52]\\
\hline
[61,70] &~\cite{54}\\
\hline
[101,110] &~\cite{39}\\
\hline
[151,160] &~\cite{41}\\
\hline
[361,370] &~\cite{45}\\
\hline
[801,810] &~\cite{53}\\
\hline
\end{tabular}
\label{RQ6}
\end{center}
\end{table}

\begin{figure}[htbp]
\centerline{\includegraphics[width=9cm, height=5cm]{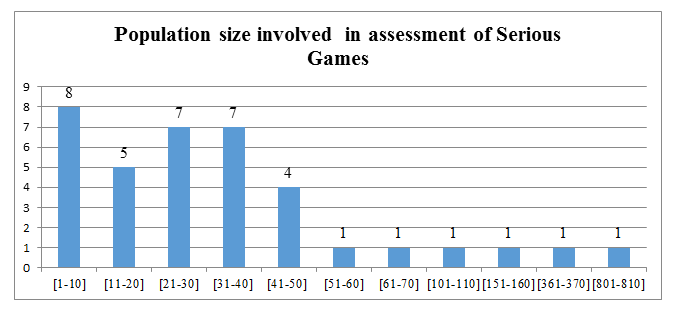}}
\caption{Distribution of population size range}
\label{fig}
\end{figure}

\begin{longtable}{| p{1cm} | p{2cm}| p{3.5cm} | p{5cm} | p{3cm} |}
\caption{Gap between previous SLR and current research}\\
\hline
\textbf{Reference}& \textbf{Online Database Name} & \textbf{Problem Discussion} & \textbf{Achievements} & \textbf{Limitations}  \\ 
\hline
\endhead
\cite{12} & Google Scholar, IEEE Xplore,  ACM Digital Library, ScienceDirect,  Springer Link  & \justifying{The paper addressed dementia disease and its symptoms (Mild Cognitive impairment (MCI) and Alzheimer’s disease (AD)). And later on, this paper gathered serious games related to dementia disease and its symptoms, i.e. Wii sports, Wii fit, Big Brain, and Academy Lumosity.} & \justifying{This review listed those serious games related to dementia disease and its symptoms, i.e. Mild Cognitive impairment (MCI) and Alzheimer’s disease (AD). After that, the paper concluded that serious games related to dementia affect cognitively impaired people.} & The only two Symptoms of dementia disease ( Mild Cognitive impairment (MCI) and Alzheimer’s disease (AD)) have been discussed, and gather only those serious games related to that disease.\\
\hline
\cite{13} & Web of Science, Embase, Cochrane Library, Psych Info, and PubMed Science.& In this systemic review, collect those papers related to serious games that can educate the public about asthma and evaluate their result on clinical outcomes, behavior, and patients' knowledge. & This paper includes those research articles on serious games. All extracted serious games were designed to educate asthma patients and the general public. Furthermore, extracted 12 articles and a total of ten serious games. All the games were related to children with asthma. Moreover, the article evaluates the result of clinical outcomes, behavior, and patients’ knowledge linked with asthma. & The limitation of this research is that they include only those articles and extract those serious games dealing with asthma patients related to children and educate the general public based on this result. Second, the study includes only 12 articles and ten serious games, which is significantly fewer in numbers. \\
\hline
\cite{14} & IEEE-Xplore, ACM Library, and ISI Web of Knowledge &This paper has performed the usability evaluation method of serious games.& This paper gathered all possible usability evaluation methods to assess serious games (Video games). This research concludes that the researcher covers those serious games used for academic purposes and evaluates the usability of serious games. As a result, they do not include games for industrial experts. In the future, we will add more demographics to the current results to show a better picture of usability evaluation in serious games.  & The main Limitation of this research has been assessed quality attributes (Usability) in serious games. And only covers serious games type video games. Research should include more serious game types like mobile games and computer games to test usability evaluation. Also, have one or two more quality attributes, like aesthetics, used in serious games.\\
\hline
\cite{15} & BioMed Central, ACM, EBSCO, SocINDEX, information technology and social
science, ERIC, Emerald and IEEE, Infotrac, Cambridge Journals Online& The problem discussed in this paper was about the positive impact of serious games on the age group of 14 years and above. For this purpose, check these aspects, i.e. skills, learning, engagement, and enhancement, and analyze the positive outcomes of mentioned aspects through serious games and computer games.  & The achievement of this research was the serious game type of computer games impact player knowledge accusation, motivational outcomes, and content understanding. It's a great achievement. The participant in the game is above 14, which means teenagers and most undergraduate students player of the game. And these outcomes were good enough according to the player's age and current knowledge. & The main limitation of this research is that it is mostly focused on serious games type computer games. Nowadays, most serious games shift towards mobile games because a player can easily carry the device without paying extra money to play the game. So, the future investigates this problem by changing the serious game type.\\
\hline
\cite{16} &  Cochrane, PubMed, EMBASE & This research examined the serious game's impact on health care training & Further investigating, the research gathered 48 serious games from the literature. In 48 serious games, 42 were unique. The serious games answered five research questions. Healthcare field, serious game name, learning goals, Genre of game, the technical resources used in the game, and availability of the game means online or offline. The research covered a vast area of health care like Surgery, Radiology, Nursing, urology, Pediatrics, etc. The second research question found the research game's name, and the third mentioned the learning goals of the games. The answer to the third research question depended on the field of the game. The fourth question listed the technical resources or technology used in serious games. The 5th question checks the availability of the game online or offline. The research was extensive and covered most of the healthcare fields. The most important thing is that the research question was very well designed.  & Moreover, the research covered a vast area of serious games and assessment of serious games. In every research area, there is a chance for improvement. This research covers mainly health care serious games type computer games. Future research will include serious mobile games because user trends have changed. Players want to avoid carrying an extra device to play a game.\\
\hline
\cite{57} & Web of Sciences, Targeted list journal and conference, Arts and 93 Humanities Citation Index (A and HCI), Science Citation Index- Expanded (SCI-Expanded) 94 and Social Science Citation Index (SSCI). & The study performed a systematic review of the literature about serious games 91 usages in science education from 2016 to 2020. & The achievement of this research was they covered multiple areas of science education. The result showed that the most important area is experimental science. This paper also covered many science subjects, i.e. Biology, Computer Science, Chemistry, STEM  and etc.Learning academic achievements was the most investigated topic. The sampling size used in this paper was 30-100 people, mostly (5-8th grade) and (9-12 grade) students. Most papers used the quantitative research design, and a descriptive analysis approach has been used to evaluate serious games. Most studies have used serious games type computer games for science education. & The limitation was this research is that it did not cover advanced areas. The future will include more advanced areas and again perform SLR to 2022 or 23 or see the latest trends of the current era in science education  \\
\hline
\cite{11} & SpringerLink, IEEE Xplore, ACM Digital Library, SCOPUS, and ISI Web of Science. & This study performed a systematic literature review of the assessment of serious games in different domains. & The achievement of this research was it covered all serious game types and domains and used all possible evaluation methods to assess serious games. For this purpose, study covered 102 papers from the preliminary study and answered the six mentioned research questions. & There is no such limitation for this research. The main gap in this SLR was the time duration they covered from 2011 to  March 2015. So there is a need to update the result. In our research, we filled these gaps and covered the 2015 onwards papers. And reinvestigate the facts and figures which have been previously outdated.\\
\hline
\end{longtable}

\label{RQ8}
\section{Discussion}
This article is about serious game evaluation and gathers articles about serious games. Serious games designed and developed for educational purposes. We performed an updated systematic literature review in the article, gathered all those serious games related to different domains, and answered the above-mentioned research questions. Furthermore, based on these research questions, we have mentioned the healthy discussion in the below table.

\begin{longtable}{| p{1.5cm} | p{7cm}| p{4.5cm} |}
\caption{Discussion Table}\\
\hline
\textbf{Research Questions}& \textbf{Discussion} & \textbf{Suggestion} \\ 
\hline
\endhead
RQ1 & In this article, primarily serious games lie in the education domain. Serious games assist different levels of students and mainly cover areas like (dental undergraduate students, Digital forensic training, Job search Skill training, and Object-oriented programming concept). Second most serious game developed for health and wellness. In this category, serious games educate people like doctors, medical students, nursing staff, and patients covering different medical areas and diseases. The most common areas covered in this review are Hemiparetic stroke patients, Cardiopulmonary Resuscitation, Laboratory testing sessions, and physical rehabilitation. Also covered were older adults' processing speed, working memory, short-term memory,  and eye-hand coordination significantly declined, phonological disorders, Autism disorder in children, Dementia disease, Dyscalculia (difficulties in learning mathematics), and Dyscalculia (challenges in learning mathematics). Serous games provide a lot of help to educate people or improve people's health, discussed in above mention diseases. Some serious games have been developed for the cultural and social domain. Covered every common issue like preventing students from bullying and cross-culture domain. Serious games help people educate about these issues and resolve their problems in minimal time. & In Rq1, our opinion is to cover more domains or investigate more serious games. Secondly, there is no such platform where anyone can be aware of serious games. How can we develop this? What is the key difference between serious games and regular games? There is a platform exists where anyone can record feedback to improve game aesthetics.\\
\hline
RQ2 & In RQ2, investigated serious game types in an updated review. Most serious games lie in the virtual worlds(3-D) based games category. Before 2016 mostly serious games were developed in computer games. But in computer games, learners are not fully satisfied with understanding complete knowledge of the required problem. While in 3-d games, beginners or intermediate-level people easily understand their respective domains.3-d Games provide a better view of any solution and present a 3-dimensional view. Most serious games are developed in 3-d and use some advanced technology like augmented reality or unity framework. Some serious games which we have investigated in our research are "Lavie Game", "A.C.R.M. Serious Games", and "Infinite Runner." Primarily 3-d serious games are used in health to educate people, doctors, nurses, and patients. This research also identified different types of serious games (Video games, Mobile games, and web-based games) & The updated systematic literature review result shows that the most serious games type virtual world(3-d).In 2022 trends have changed. Users prefer to play games on their mobile phones. Hardly people use a computer nowadays. Yes, computer science and experts still use laptops or desktop pcs for their heavy computational. But my advice is to develop serious games for mobile and 3-d. Because serious game type mobile games have a lot of benefits, firstly they can easily carry a device,2nd no additional cost to play a game. The key benefit is they can educate anywhere through the serious game.\\
\hline
RQ3 & In RQ3 investigated the evaluation method of serious games. The evaluation methods questionnaire and frameworks assess mostly serious games. The types of questionnaires and frameworks have been different according to domain types. The serious games have been developed for educational purposes and used questionnaires. The type of questionnaires depends on which area of education have covered serious games. Other than these two evaluation methods, some more assessments have been used to evaluate the serious games, i.e. Evaluation and training, experimental, interviews, Thermography, cognitive skills, user experience, audio-video feedback, and observation were also used to evaluate the serious games. & There are a lot of methods we have identified to evaluate serious games. In my opinion, only one evaluation method exists for a specific domain. The benefit is somebody develops a serious game in that domain. They choose the same evaluation method, which saves them a lot of time.\\
\hline
RQ4 & RQ4 have identified 13 quality attributes to evaluate serious games. In most serious games, usability quality attributes have been used to assess serious games. In most serious games, people don't know how to play the game. or how to change options according to their needs. If the player hard to learn the game's user interface, they leave that game. Usability quality attributes are mainly used in education and health-related serious games. Secondly, mostly serious games have been evaluated with the quality attribute "performance" if the serious game performance is not up to mark or slow, the player hard to learn their problem and invests a lot of time to learn the solution to a particular situation. Other attributes that have been used to evaluate serious games in this research, i.e. Game design, learning outcomes, User interface, User satisfaction, Cognitive behavior, Learning Efficacy, Learnability, Pedagogical aspects and Playability. & As a result of current and previous SLR results, many quality attributes have been used to evaluate serious games. It's only the developer's choice which quality attributes they choose for developing serious games. Nowadays, playability and usability are the key quality attribute to evaluate a serious game. Players can easily understand how to play and operate the game using this quality attribute. In most serious games developed for mobile platforms, the quality attribute usability is a key attribute.\\
\hline
RQ5 & RQ5 searched out the evaluation method of serious games and analyzed how to assess serious games.Pre/post-tests have been commonly used to analysis of serious games. This evaluation method collects the result before and after playing the game. This method quickly assesses the player's performance or learning through the game. Statistical methods have also been used to evaluate serious games. Some statistical methods are Z-test, T-test, Mann-Whitney TESTS, and Shapiro–Wilk test. The experimental test has also been used to analysis of serious games. & In RQ5, we have searched for serious game evaluation methods. Analyze how methods have been evaluated in serious games. Mostly serious games used Pre/post evaluation method to assess the serious games. This method is suitable because it gives a clear picture of learning in front of the player before and after playing the serious game. In my opinion, one generalized way for a specific domain is to check the evolution of serious games.\\
\hline
RQ6 & RQ6 evaluated the sample size of serious games. In most of the serious games, the sample size was composed of 1 to 10 people. Some of the serious game's samples size was above 100. The overall sample size depended on the nature of the game. If the collected sample and population size are correct, then defiantly affects the game's accuracy. & This question identified the population size to evaluate the serious games. Population sizes of 1 to 10 people have mostly been used to assess serious games. In fewer serious games, the population size is above 100 people. In my opinion, fix the size of the population according to the domain of serious games.\\
\hline
\end{longtable}
\label{RQ1}

\section{Conclusion}
This study found 37 papers demonstrating the techniques, methods, and approaches used to evaluate serious games. We structured, listed, and analyzed the information we received to answer the six research questions discussed in this study. We performed the experiences to evaluate the various types of serious games in different application domains. Then, we identified the different methods, techniques, and approaches to assess serious games. Our updated SLR will help future researchers and practitioners evaluate serious games and provide linkage with previous studies. Following is the summary response to each research question.

\begin{itemize}
    \item Answer to RQ1: In this study, we concluded 37 papers that explored six application domains i.e. culture, education, health and wellness, support, professional learning and training, and social. In the education domain, the highest numbers of serious games have been evaluated.
    \item Answer of RQ2: In 37 papers, seven types of serious games have been identified i.e. computer games, mobile games, virtual worlds, Lego-based games, board games, and web-based games. Most domains had used virtual worlds 3D games, which were the highest serious games type. The virtual world 3D games were frequently used for medical and educational purposes to train and assist people.
    \item Answer of RQ3: In 37 papers, we identified 11 assessment methods as discussed above. Most studies used questionnaires and frameworks to evaluate serious games.
    \item Answer of RQ4: In this study, 13 quality attributes had been identified in our 37 selected papers. Most studies used usability as a quality attribute to evaluate serious games.
    \item Answer of RQ5:In the response, 17 assessment methods were identified to evaluate the serious games. Some statistical methods were also used like the Shapiro-Wilk test, post-hoc tests, Maan-Whitney test, and T-test for the evaluation purpose.
    \item Answer of RQ6: In these questions identified population sizes that had been used to evaluate the serious games. In 37 papers, most studies used participants' size range of no more than 50 subjects to evaluate serious games.
\end{itemize}

our updated SLR provides a comprehensive overview of the techniques, methods, and approaches used to evaluate serious games. The findings from this study will serve as a valuable resource for future researchers and practitioners in the field, enabling them to build upon existing knowledge and establish connections with previous studies. By enhancing the evaluation of serious games, we can further advance their development and impact in various domains.

\bibliographystyle{unsrt}
\bibliography{SeriousGames}   

\begin{thebibliography}{10}

\bibitem{1}
Charles~D Elder.
\newblock Serious games. by clark c. abt.(new york: The viking press, inc.,
  1970. pp. 176. 4.95 paper.).
\newblock {\em American Political Science Review}, 65(4):1158--1159, 1971.

\bibitem{2}
Meng-Tzu Cheng, Jhih-Hao Chen, Sheng-Ju Chu, and Shin-Yen Chen.
\newblock The use of serious games in science education: a review of selected
  empirical research from 2002 to 2013.
\newblock {\em Journal of computers in education}, 2(3):353--375, 2015.

\bibitem{8}
Marc-Andr{\'e} Maheu-Cadotte, Sylvie Cossette, V{\'e}ronique Dub{\'e},
  Guillaume Fontaine, Tanya Mailhot, Patrick Lavoie, Alexis Cournoyer, Fabio
  Balli, and Gabrielle Mathieu-Dupuis.
\newblock Effectiveness of serious games and impact of design elements on
  engagement and educational outcomes in healthcare professionals and students:
  a systematic review and meta-analysis protocol.
\newblock {\em BMJ open}, 8(3):e019871, 2018.

\bibitem{11}
Alejandro Calder{\'o}n and Mercedes Ruiz.
\newblock A systematic literature review on serious games evaluation: An
  application to software project management.
\newblock {\em Computers \& Education}, 87:396--422, 2015.

\bibitem{12}
Simon McCallum and Costas Boletsis.
\newblock Dementia games: A literature review of dementia-related serious
  games.
\newblock In {\em International conference on serious games development and
  applications}, pages 15--27. Springer, 2013.

\bibitem{13}
David Drummond, Delphine Monnier, Antoine Tesni{\`e}re, and Alice Hadchouel.
\newblock A systematic review of serious games in asthma education.
\newblock {\em Pediatric Allergy and Immunology}, 28(3):257--265, 2017.

\bibitem{14}
Rosa Yanez-Gomez, Daniel Cascado-Caballero, and Jose-Luis Sevillano.
\newblock Academic methods for usability evaluation of serious games: a
  systematic review.
\newblock {\em Multimedia Tools and Applications}, 76(4):5755--5784, 2017.

\bibitem{15}
Thomas~M Connolly, Elizabeth~A Boyle, Ewan MacArthur, Thomas Hainey, and
  James~M Boyle.
\newblock A systematic literature review of empirical evidence on computer
  games and serious games.
\newblock {\em Computers \& education}, 59(2):661--686, 2012.

\bibitem{16}
Ryan Wang, Samuel DeMaria~Jr, Andrew Goldberg, and Daniel Katz.
\newblock A systematic review of serious games in training health care
  professionals.
\newblock {\em Simulation in Healthcare}, 11(1):41--51, 2016.

\bibitem{57}
Nuri Kara.
\newblock A systematic review of the use of serious games in science education.
\newblock {\em Contemporary Educational Technology}, 13(2):ep295, 2021.

\bibitem{58}
Thiago~Machado Leit{\~a}o, Leonardo Luiz~Lima Navarro, Renato~Fl{\'o}rido
  Cameira, and Edison~Renato Silva.
\newblock Serious games in business process management: a systematic literature
  review.
\newblock {\em Business Process Management Journal}, 2021.

\bibitem{10}
Staffs Keele et~al.
\newblock Guidelines for performing systematic literature reviews in software
  engineering.
\newblock Technical report, Citeseer, 2007.

\bibitem{18}
Barbara Kitchenham and Stuart Charters.
\newblock Guidelines for performing systematic literature reviews in software
  engineering.
\newblock 2007.

\bibitem{19}
Brian An, Donald Brown, and Stephanie Guerlain.
\newblock The evaluation of a serious game to improve cross-cultural
  competence.
\newblock {\em IEEE Transactions on Learning Technologies}, 12(3):429--441,
  2019.

\bibitem{20}
F~Noveletto, AV~Soares, BA~Mello, CN~Sevegnani, FLF Eichinger, M~da~S Hounsell,
  and P~Bertemes-Filho.
\newblock Biomedical serious game system for balance rehabilitation of
  hemiparetic stroke patients.
\newblock {\em IEEE Transactions on Neural Systems and Rehabilitation
  Engineering}, 26(11):2179--2188, 2018.

\bibitem{22}
Muhammad~Hamza Latif, Muhammad Ajmal, Farooq Ahmad, Junaid Alam, and Asma
  Saleem.
\newblock La-vie: A serious game for cardiopulmonary resuscitation.
\newblock In {\em 2017 IEEE 5th International Conference on Serious Games and
  Applications for Health (SeGAH)}, pages 1--5. IEEE, 2017.

\bibitem{29}
Octavian Postolache, Filipe Louren{\c{c}}o, JM~Dias Pereira, and Pedro
  Gir{\~a}o.
\newblock Serious game for physical rehabilitation: Measuring the effectiveness
  of virtual and real training environments.
\newblock In {\em 2017 IEEE International Instrumentation and Measurement
  Technology Conference (I2MTC)}, pages 1--6. IEEE, 2017.

\bibitem{30}
Diogo Ferreira, Raul Oliveira, and Octavian Postolache.
\newblock Physical rehabilitation based on kinect serious games.
\newblock In {\em 2017 Eleventh international conference on sensing technology
  (ICST)}, pages 1--6. IEEE, 2017.

\bibitem{31}
Yu-Hsiang Lin, Hui-Fen Mao, Yun-Cheng Tsai, and Jui-Jen Chou.
\newblock Developing a serious game for the elderly to do physical and
  cognitive hybrid activities.
\newblock In {\em 2018 IEEE 6th International Conference on Serious Games and
  Applications for Health (SeGAH)}, pages 1--8. IEEE, 2018.

\bibitem{32}
Bonnech{\`e}re Bruno, Berlemont Christophe, Feipel V{\'e}ronique, Jansen Bart,
  et~al.
\newblock A preliminary study of the integration of specially developed serious
  games in the treatment of hospitalized elderly patients.
\newblock In {\em 2017 International Conference on Virtual Rehabilitation
  (ICVR)}, pages 1--6. IEEE, 2017.

\bibitem{37}
Rui~Neves Madeira, Vanessa Mestre, and T{\^a}nia Ferreirinha.
\newblock Phonological disorders in children? design and user experience
  evaluation of a mobile serious game approach.
\newblock {\em Procedia computer science}, 113:416--421, 2017.

\bibitem{40}
Kamran Khowaja and Siti~Salwah Salim.
\newblock Serious game for children with autism to learn vocabulary: An
  experimental evaluation.
\newblock {\em International journal of human--computer interaction},
  35(1):1--26, 2019.

\bibitem{42}
Rytis Maskeli{\=u}nas, Robertas Dama{\v{s}}evi{\v{c}}ius, Tomas
  Bla{\v{z}}auskas, Andrius Paulauskas, Lukas Paulauskas, and C~Chiatti.
\newblock Ido: modelling a serious educational game based on hands on approach
  for training dementia carers.
\newblock {\em International Journal of Engineering and Technology},
  7(2.28):143--146, 2018.

\bibitem{47}
Diego~Fernando Avila-Pesantez, Leticia~Azucena Vaca-Cardenas, Rosa~Delgadillo
  Avila, Nelly~Padilla Padilla, and Luis~A Rivera.
\newblock Design of an augmented reality serious game for children with
  dyscalculia: a case study.
\newblock In {\em International Conference on Technology Trends}, pages
  165--175. Springer, 2018.

\bibitem{48}
Costas Boletsis and Simon McCallum.
\newblock Augmented reality cubes for cognitive gaming: preliminary usability
  and game experience testing.
\newblock 2016.

\bibitem{21}
Rafael~Oliveira Chaves, Christiane~Gresse von Wangenheim, Julio Cezar~Costa
  Furtado, Sandro Ronaldo~Bezerra Oliveira, Alex Santos, and Eloi~Luiz Favero.
\newblock Experimental evaluation of a serious game for teaching software
  process modeling.
\newblock {\em ieee Transactions on Education}, 58(4):289--296, 2015.

\bibitem{23}
Abdelbaset~Jamal Abdellatif, Barry McCollum, and Paul McMullan.
\newblock Serious games: Quality characteristics evaluation framework and case
  study.
\newblock In {\em 2018 IEEE Integrated STEM Education Conference (ISEC)}, pages
  112--119. IEEE, 2018.

\bibitem{24}
Ivana Gace, Lucija Jaksic, Ilir Murati, Iva Topolovac, Matea Zilak, and Zeljka
  Car.
\newblock Virtual reality serious game prototype for presenting military units.
\newblock In {\em 2019 15th International Conference on Telecommunications
  (ConTEL)}, pages 1--6. IEEE, 2019.

\bibitem{25}
Ugyen Nima, Rinzin Wangdi, and Jannicke~Baalsrud Hauge.
\newblock A serious game for competence development in internet of things and
  knowledge sharing.
\newblock In {\em 2018 IEEE International Conference on Industrial Engineering
  and Engineering Management (IEEM)}, pages 1786--1790. IEEE, 2018.

\bibitem{26}
Tobias Mettler and Roberto Pinto.
\newblock Serious games as a means for scientific knowledge transfer—a case
  from engineering management education.
\newblock {\em IEEE Transactions on Engineering Management}, 62(2):256--265,
  2015.

\bibitem{27}
Maria Drakou and Andreas Lanitis.
\newblock On the development and evaluation of a serious game for forensic
  examination training.
\newblock In {\em 2016 18th Mediterranean Electrotechnical Conference
  (MELECON)}, pages 1--6. IEEE, 2016.

\bibitem{28}
Justin~B Marshall, Gary Tyson, Juan Llanos, Roberto~Miguel Sanchez, and
  Francisca~B Marshall.
\newblock Serious 3d gaming research for the vision impaired.
\newblock In {\em 2015 17th International Conference on E-health Networking,
  Application \& Services (HealthCom)}, pages 468--471. IEEE, 2015.

\bibitem{33}
Hege~Mari Johnsen, Mariann Fossum, Pirashanthie Vivekananda-Schmidt, Ann
  Fruhling, and {\AA}shild Sletteb{\o}.
\newblock Teaching clinical reasoning and decision-making skills to nursing
  students: Design, development, and usability evaluation of a serious game.
\newblock {\em International journal of medical informatics}, 94:39--48, 2016.

\bibitem{34}
Jungmin Kwon and Youngsun Lee.
\newblock Serious games for the job training of persons with developmental
  disabilities.
\newblock {\em Computers \& Education}, 95:328--339, 2016.

\bibitem{35}
Omair Ameerbakhsh, Savi Maharaj, Amir Hussain, and Bruce McAdam.
\newblock A comparison of two methods of using a serious game for teaching
  marine ecology in a university setting.
\newblock {\em International journal of human-computer studies}, 127:181--189,
  2019.

\bibitem{36}
Harits~Ar Rosyid, Matt Palmerlee, and Ke~Chen.
\newblock Deploying learning materials to game content for serious education
  game development: A case study.
\newblock {\em Entertainment computing}, 26:1--9, 2018.

\bibitem{41}
Kawin Sipiyaruk, Jennifer~E Gallagher, Stylianos Hatzipanagos, and Patricia~A
  Reynolds.
\newblock Acquiring critical thinking and decision-making skills: An evaluation
  of a serious game used by undergraduate dental students in dental public
  health.
\newblock {\em Technology, Knowledge and Learning}, 22(2):209--218, 2017.

\bibitem{43}
Andrej~Jerman Bla{\v{z}}i{\v{c}}, Primo{\v{z}} Cigoj, and Borka~Jerman
  Bla{\v{z}}i{\v{c}}.
\newblock Serious game design for digital forensics training.
\newblock In {\em 2016 Third International Conference on Digital Information
  Processing, Data Mining, and Wireless Communications (DIPDMWC)}, pages
  211--215. IEEE, 2016.

\bibitem{44}
Ufuk Aydan, Murat Yilmaz, Paul~M Clarke, and Rory~V O’Connor.
\newblock Teaching iso/iec 12207 software lifecycle processes: A serious game
  approach.
\newblock {\em Computer Standards \& Interfaces}, 54:129--138, 2017.

\bibitem{45}
CM~Steiner, Alexander Nussbaumer, Kerstin Gaisbachgrabner, Thierry Platon,
  Olivier Lepoivre, and Dietrich Albert.
\newblock Evaluation of an interview simulation for job seekers-applying a
  holistic evaluation approach for serious games.
\newblock In {\em 13th annual International Technology, Education and
  Development Conference}, 2018.

\bibitem{49}
M~J Callaghan, Niall McShane, A~Gomez Eguiluz, T~Teilles, and P~Raspail.
\newblock Practical application of the learning mechanics-game mechanics
  (lm-gm) framework for serious games analysis in engineering education.
\newblock In {\em 2016 13th International Conference on Remote Engineering and
  Virtual Instrumentation (REV)}, pages 391--395. IEEE, 2016.

\bibitem{51}
Robert Shewaga, Alvaro Uribe-Quevedo, Bill Kapralos, Kenneth Lee, and Fahad
  Alam.
\newblock A serious game for anesthesia-based crisis resource management
  training.
\newblock {\em Computers in Entertainment (CIE)}, 16(2):1--16, 2018.

\bibitem{52}
Simone Ghisio, Paolo Alborno, Erica Volta, Monica Gori, and Gualtiero Volpe.
\newblock A multimodal serious-game to teach fractions in primary school.
\newblock In {\em Proceedings of the 1st ACM SIGCHI International Workshop on
  Multimodal Interaction for Education}, pages 67--70, 2017.

\bibitem{53}
Gokul CJ, Sankalp Pandit, Sukanya Vaddepalli, Harshal Tupsamudre, Vijayanand
  Banahatti, and Sachin Lodha.
\newblock Phishy-a serious game to train enterprise users on phishing
  awareness.
\newblock In {\em Proceedings of the 2018 annual symposium on computer-human
  interaction in play companion extended abstracts}, pages 169--181, 2018.

\bibitem{54}
Alexis~D Souchet, St{\'e}phanie Philippe, Dimitri Zobel, Floriane Ober,
  Aur{\'e}lien L{\'e}v{\v{e}}que, and Laure Leroy.
\newblock Eyestrain impacts on learning job interview with a serious game in
  virtual reality: a randomized double-blinded study.
\newblock In {\em Proceedings of the 24th ACM Symposium on Virtual Reality
  Software and Technology}, pages 1--12, 2018.

\bibitem{55}
{\v{C}}en{\v{e}}k {\v{S}}a{\v{s}}inka, Zden{\v{e}}k Stacho{\v{n}}, Michal
  Sedl{\'a}k, Ji{\v{r}}{\'\i} Chmel{\'\i}k, Luk{\'a}{\v{s}} Herman, Petr
  Kub{\'\i}{\v{c}}ek, Al{\v{z}}b{\v{e}}ta {\v{S}}a{\v{s}}inkov{\'a}, Milan
  Dole{\v{z}}al, Hynek Tejkl, Tom{\'a}{\v{s}} Urb{\'a}nek, et~al.
\newblock Collaborative immersive virtual environments for education in
  geography.
\newblock {\em ISPRS International Journal of Geo-Information}, 8(1):3, 2019.

\bibitem{38}
David~W Wilson, Jeff Jenkins, Nathan Twyman, Matthew Jensen, Joe Valacich,
  Norah Dunbar, Scott Wilson, Claude Miller, Bradly Adame, Yu-Hao Lee, et~al.
\newblock Serious games: an evaluation framework and case study.
\newblock In {\em 2016 49th Hawaii International Conference on System Sciences
  (HICSS)}, pages 638--647. IEEE, 2016.

\bibitem{39}
David~W Wilson, Jeff Jenkins, Nathan Twyman, Matthew Jensen, Joe Valacich,
  Norah Dunbar, Scott Wilson, Claude Miller, Bradly Adame, Yu-Hao Lee, et~al.
\newblock Serious games: an evaluation framework and case study.
\newblock In {\em 2016 49th Hawaii International Conference on System Sciences
  (HICSS)}, pages 638--647. IEEE, 2016.

\bibitem{46}
Katie Li, Mark Hall, Pablo Bermell-Garcia, Jeffrey Alcock, Ashutosh Tiwari, and
  Mar Gonz{\'a}lez-Franco.
\newblock Measuring the learning effectiveness of serious gaming for training
  of complex manufacturing tasks.
\newblock {\em Simulation \& Gaming}, 48(6):770--790, 2017.

\bibitem{50}
C{\'a}tia Raminhos, Ana~Paula Cl{\'a}udio, Maria~Beatriz Carmo, Augusta Gaspar,
  Susana Carvalhosa, and Maria de~Jesus~Candeias.
\newblock A serious game-based solution to prevent bullying.
\newblock {\em International Journal of Pervasive Computing and
  Communications}, 2016.

\end{thebibliography}

\end{document}